anonymous
\documentstyle[12pt,epsf]{article}
\begin{document}
\catcode`@=11
% Redefine caption to put text and formulas in smaller font
\long\def\@caption#1[#2]#3{\par\addcontentsline{\csname
  ext@#1\endcsname}{#1}{\protect\numberline{\csname
  the#1\endcsname}{\ignorespaces #2}}\begingroup
    \small
    \@parboxrestore
    \@makecaption{\csname fnum@#1\endcsname}{\ignorespaces #3}\par
  \endgroup}
\catcode`@=12
\def\marginnote#1{}
%%%%%%%%%%%%%%%%%% defs %%%%%%%%%%%%%%%%%%
\newcommand{\newc}{\newcommand}
\newc{\mtop}{\mt}
\newc{\btau}{$b$--$\tau$}
\newc{\bino}{\widetilde B}
\newc{\wino}{\widetilde W}
\newc{\beq}{\begin{equation}}
\newc{\eeq}{\end{equation}}
\newc{\bea}{\begin{eqnarray}}
\newc{\eea}{\end{eqnarray}}
\newc{\onehalf}{\frac{1}{2}}
\newc{\gsim}{\lower.7ex\hbox{$\;\stackrel{\textstyle>}{\sim}\;$}}
\newc{\lsim}{\lower.7ex\hbox{$\;\stackrel{\textstyle<}{\sim}\;$}}
\newc{\alphas}{\alpha_s}
\newc{\tanb}{\tan\beta}
\newc{\mz}{m_Z}		\newc{\mw}{m_W}
\newc{\mhalf}{m_{1/2}}
\newc{\mzero}{m_0}
\newc{\muzero}{\mu_0}
\newc{\sgnmu}{{\rm sgn}\,\mu}
\newc{\azero}{A_0}
\newc{\Atop}{A_t}
\newc{\bzero}{B_0}
\newc{\mt}{m_t}
\newc{\mb}{m_b}
\newc{\mtau}{m_\tau}
\newc{\mq}{m_q}
\newc{\htop}{h_t}
\newc{\hbot}{h_b}
\newc{\htau}{h_\tau}
\newc{\mtpole}{M_t}
\newc{\mbpole}{M_b}
\newc{\mqpole}{M_q}
\newc{\mgut}{M_X}
\newc{\mx}{\mgut}
\newc{\alphax}{\alpha_X}
\newc{\ie}{{\it i.e.}}
\newc{\etal}{{\it et al.}}
\newc{\eg}{{\it e.g.}}
\newc{\etc}{{\it etc.}}
\newc{\hh}{{h^0}}
\newc{\mhh}{m_\hh}
\newc{\hH}{{H^0}}
\newc{\mhH}{m_\hH}
\newc{\hA}{{A^0}}
\newc{\mhA}{m_\hA}
\newc{\hpm}{{H^\pm}}
\newc{\mhpm}{m_\hpm}
\newc{\stp}{{\widetilde t}}
\newc{\stopl}{{\stp_L}}
\newc{\stopr}{{\stp_R}}

\newc{\stau}{{\widetilde\tau}}
\newc{\stauone}{\widetilde\tau_1}
\newc{\mstauone}{m_\stauone}

\newc{\gluino}{{\widetilde g}}	\newc{\mgluino}{m_{\gluino}}
\newc{\MS}{{\rm\overline{MS}}}
\newc{\DR}{{\rm\overline{DR}}}
\newc{\ev}{{\rm\,eV}}
\newc{\gev}{{\rm\,GeV}}
\newc{\tev}{{\rm\,TeV}}
\newc{\bsg}{BR$(b\to s\gamma)$}
\newc{\abundchi}{\Omega_\chi h_0^2}

\newc{\mchi}{m_\chi}
\newc{\mcharone}{m_{\charone}}	\newc{\charone}{\chi_1^\pm}
\newc{\mneutone}{m_{\neutone}}	\newc{\neutone}{\chi^0_1}
\newc{\mneuttwo}{m_{\neuttwo}}	\newc{\neuttwo}{\chi^0_2}
\def\NPB#1#2#3{Nucl. Phys. {\bf B#1} (19#2) #3}
\def\PLB#1#2#3{Phys. Lett. {\bf B#1} (19#2) #3}
\def\PLBold#1#2#3{Phys. Lett. {\bf#1B} (19#2) #3}
\def\PRD#1#2#3{Phys. Rev. {\bf D#1} (19#2) #3}
\def\PRL#1#2#3{Phys. Rev. Lett. {\bf#1} (19#2) #3}
\def\PRT#1#2#3{Phys. Rep. {\bf#1} (19#2) #3}
\def\ARAA#1#2#3{Ann. Rev. Astron. Astrophys. {\bf#1} (19#2) #3}
\def\ARNP#1#2#3{Ann. Rev. Nucl. Part. Sci. {\bf#1} (19#2) #3}
\def\MODA#1#2#3{Mod. Phys. Lett. {\bf A#1} (19#2) #3}
\def\ZPC#1#2#3{Zeit. f\"ur Physik {\bf C#1} (19#2) #3}
\def\APJ#1#2#3{Ap. J. {\bf#1} (19#2) #3}
%%%%%%%%%%%%% end of defs %%%%%%%%%%%%%%%
\begin{titlepage}
\begin{flushright}
{\large
UM-TH-94-18\\
hep-ph/9405364\\
April 1994\\
}
\end{flushright}
\vskip 2cm
\begin{center}
{\large\bf UPPER BOUNDS IN LOW-ENERGY SUSY}
\vskip 1cm
{\large
G.L. Kane,
Chris Kolda,
Leszek Roszkowski\footnote{
Talk at the 2nd IFT Workshop {\it Yukawa Couplings and the Origin of
Mass},
Gainesville, Florida, February 11--13, 1994.},
and James D. Wells
\\}
\vskip 2pt
{\it Randall Physics Laboratory, University of Michigan,\\ Ann Arbor,
MI 48190, USA}\\
\end{center}
\vskip 1.5cm
\begin{abstract}
In the constrained MSSM one is typically able to restrict the
supersymmetric
mass spectra
below roughly 1-2\tev\ {\em without} resorting to  the ambiguous
fine-tuning
constraint.
\end{abstract}
\end{titlepage}
\setcounter{footnote}{0}
\setcounter{page}{2}
\setcounter{section}{0}
%\newpage

\section{Constrained MSSM (CMSSM)}

Low-energy SUSY has been considered an attractive extension of the
Standard
Model (SM) ever since it was introduced over a decade
ago.  In the early days many expected supersymmetric masses to lie
rather low,
``just around the corner", often well within the reach of LEP and the
Tevatron. Not finding SUSY signals there was consequently rather
disappointing and one could hear from sceptics sarcastic comments
that
a SUSY discovery will always remain to be expected for the {\em next}
round of
accelerators, in a time-invariant manner.
Theoretical arguments based on no fine-tuning limiting SUSY masses
roughly
below 1\tev\ were greeted with even less trust. After all,
theorists are known to be both creative and, at the same time, rather
unwilling to give up their most beloved toys, as the continuing
activity in
alternatives to SUSY clearly shows.

In this talk I am going to show that, in the Minimal Supersymmetric
Standard
Model (MSSM) with a few sensible and relatively general assumptions,
one is
often able to limit the SUSY particle masses
below about 1-2\tev\ {\em by physical constraints
alone}~\cite{kkrwone,kkrwtwo},
without having to resort to an ill-defined fine-tuning
constraint. Furthermore, the assumptions that we make are actually
typically
also
made in most
phenomenological studies of the MSSM and are well-motivated by GUTs.
I will
call this framework the constrained MSSM (CMSSM)~\cite{kkrwone}.

First, it is worth remembering that SUSY alone has been applied to
particle physics in order to provide a sensible framework for GUTs,
which are
otherwise plagued by the (in)famous problems of naturalness and scale
hierarchy. Without GUTs, or related attempts (like strings) to not
only unify
all
interactions but also to close the gap between to Fermi scale and the
only fundamental scale in high energy physics, the Planck scale,
there
is indeed little motivation to consider low-energy SUSY.
Furthermore, precision measurements at LEP have provided us with
a remarkable argument for gauge coupling unification within (even
minimal)
SUSY, while showing more than clearly that within the SM alone
such unification does not take place~\cite{early}.
We will thus require that gauge couplings unify which will, for our
purpose,
fix the unification scale
$\mx$. By doing so we actually are not forced to assume the existence
of any specific GUT. We will only assume that $\sin^2\theta_{\rm
w}(\mx)=\frac{3}{8}$ which
also holds in many phenomenologically viable superstring-derived
models.

Second, if the idea of unification is to be taken seriously, then one
should expect not only the gauge couplings to emerge from a common
source but
also
the same to be true for the various mass parameters of low-energy
SUSY. In
particular, one typically assumes that all the mass terms of the
scalars in the model, like the squarks, sleptons and the Higgs
bosons,
originate from one ``common" source $\mzero$. Similarly, the masses
of
the gauginos (the gluino, winos and bino) should be equal to the
``common"
gaugino mass $\mhalf$ at $\mx$. These two assumptions are
certainly not irrefutable but are at least sensible. In addition,
they
result from the simplest minimal supergravity framework and the
simplest choice
of the kinetic potential. Furthermore, there is at least some partial
motivation for assuming the common scalar mass $\mzero$ coming from
experiment.
The near mass
degeneracy in the $K^0-\bar K^0$ system implies
a near mass degeneracy
between $\widetilde s_L$ and $\widetilde d_L$~\cite{dgs}.
Similarly, some slepton masses have to be strongly degenerate from
stringent bounds on $\mu\rightarrow e\gamma$~\cite{dgs}.
Needless to say, most phenomenological studies of SUSY rely on at
least
one of these two assumptions, at least for the sake of reducing the
otherwise
huge number of unrelated SUSY mass parameters. We also assume
that the trilinear soft SUSY-breaking terms are equal to $\azero$ at
$\mx$,
although this assumption has actually almost no bearing here.

Furthermore, it has been long known that in SUSY there exists a
remarkable
``built-in" mechanism of radiative electroweak symmetry breaking
(EWSB). When
the Higgs mass-square parameters are run from the high scale
down, at some point the Higgs fields develop vevs. We thus require
that the conditions for EWSB be satisfied.

Having made these sensible and well-motivated assumptions, we can
next
derive complete mass spectra of all the Higgs and supersymmetric
particles by
running their 1-loop RGEs between $\mx$ and $\mz$.
The spectra are parametrized in terms of just a few basic parameters
which we
conveniently choose to be: the top mass $\mtpole$, $\tanb$, $\mhalf$,
$\mzero$, as well as $\azero$. The parameters $|\mu|$ and $B$ are
determined
through the conditions for EWSB, but the sign of $\mu$ remains
undetermined. We
thus consider both $\sgnmu=\pm1$.
We also employ the full 1-loop effective Higgs potential.

Besides requiring that EWSB occur, we demand that all physical
mass-squares remain positive. We impose mass limits from current
direct
experimental searches and include
the requirement that the solutions provide a \bsg\ consistent with
CLEO
data. Furthermore, we calculate
the relic density of the lightest SUSY particle (LSP), demanding only
that the
LSP be neutral, and, from limits on the age of the Universe of 10
billion
years, we demand that $\Omega_{LSP}h_0^2<1$.
Those solutions which finally remain after
all these cuts comprise the allowed parameter space of the CMSSM.

\section{Upper Limits}

We have explored wide ranges of parameters, as described in detail in
Refs.~\cite{kkrwone} and~\cite{kkrwtwo}. Clearly, in general one
expects the
emerging
patterns of the SUSY mass spectra and properties (mixings, \etc)
to vary strongly with the input parameters. While this is indeed true
to some extent, nevertheless certain universal features emerge.

\begin{figure}
\centering
\epsfysize=2.75in
\hspace*{0in}
\epsffile{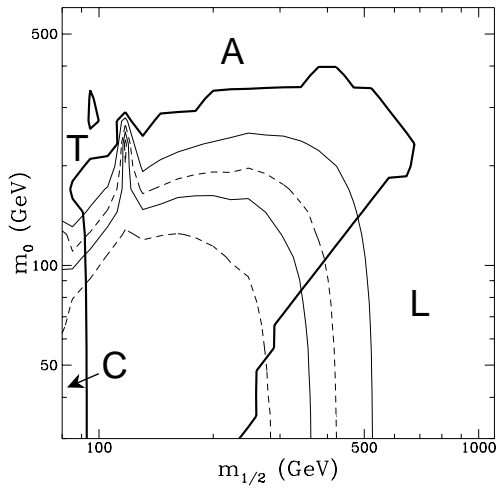}
\caption{The regions of the $(\mhalf,\mzero)$ plane consistent with
low $\tanb$ \btau\ mass unification, given all the constraints of the
CMSSM, for $\mtpole=170\gev$, $\azero/\mzero=0$ and $\mu<0$.
Solutions outside the thick solid lines are excluded: on the left
(small
$\mhalf$) by the chargino mass bound ({\bf C})
$m_{\chi^\pm}>47\gev$ and by tachyonic $\stp$'s ({\bf T}); on the
right
(large $\mhalf\gg\mzero$) by charged LSP ({\bf L}); and from above by
the age
of the Universe, \ie\ $\abundchi\leq1$ ({\bf A}). We also indicate
the
sub-regions selected by either the hypothesis of cold dark matter
($0.25\lsim\abundchi\lsim0.5$, between thin solid lines) or the one
of mixed
dark matter ($0.16\lsim\abundchi\lsim0.33$, between thin dashed
lines).
}
\label{dm:fig}
\end{figure}

\noindent These features are illustrated in Fig.~\ref{dm:fig} for
$\mtpole=170\gev$. The region of
small $\mhalf$ is always excluded by either direct experimental
searches for
SUSY at LEP (typically the strongest bounds come from chargino or
Higgs mass
limits) or
at the Tevatron (gluino). In some cases, in particular for
$|\azero|/\mzero$
significantly above zero, the lighter stop
becomes too light, and even tachyonic, for
$\mzero\gg\mhalf\lsim100\gev$.
Also, for some but rare combinations of parameters, for either
$\mzero\gg\mhalf$ or $\mhalf\gg\mzero$
the conditions for EWSB fail to be satisfied.

Furthermore, since we assume unbroken $R$-parity here, the LSP is
stable and
should be present as a relic in the Universe. There are strong
arguments
against charged exotic relics. We thus require the LSP
to be neutral. This rules out a significant region of the
($\mhalf,\mzero$)
parameter space corresponding to $\mhalf\gg\mzero$
where the LSP is the lighter stau $\stauone$. (Also $\widetilde e_R$
and
$\widetilde\mu_R$ are not much heavier there.)

In the remaining region allowed by all the conditions listed above it
is the
lightest neutralino $\chi$
that is
the LSP. (The sneutrino, another neutral sparticle, is the LSP in the
region of
small $\mhalf$ which is now completely excluded experimentally.) This
is quite
remarkable given the fact that the neutralino is a very attractive
candidate
for the dark matter (DM) in the Universe for which there seems to be
inescapable need among astrophysicists~\cite{dmreview}. Equally
remarkable
and non-trivial is the fact that
$\chi$ comes out mostly bino-like which is essentially a necessary
condition
if one expects the neutralino to be a significant component of DM in
the
Universe. (The neutralino with a significant higgsino admixture
has invariably very small relic abundance~\cite{dmreview}.)
The only exceptions to this general rule can be found in some
relatively rare
case in very tiny regions of the ($\mhalf,\mzero$) on the border of
the region where the conditions for the EWSB cannot be satisfied.

The fact that the lightest neutralino of bino-type comes out in the
CMSSM as
the unique neutral candidate for the LSP is not only interesting in
itself. It
also leads to a very remarkable upper bound
on both $\mhalf$ and $\mzero$. This comes about as follows. The
neutralino
relic density $\rho_\chi$ depends on how many neutralinos
have pair-annihilated in the early Universe. Their number effectively
froze
when the expansion rate exceeded the annihilation rate.
In order to calculate the neutralino relic density one thus needs to
include all the annihilation channels of the neutralinos into
ordinary
particles, which we do. Since all the masses and mixings are
determined
in the CMSSM in terms of the basic independent parameters listed
above,
one can also express in terms of them the neutralino relic
abundance $\abundchi$ (which is the neutralino relic density in units
of the
critical density times the squared reduced Hubble constant).
The key point is that any significant contribution to the total
mass-energy
density of the Universe would have affected its evolution.
In particular, the greater the total density the faster the Universe
expands and the more quickly it reaches its present size. The age
of the Universe, which is known to be {\em at least} 10 billion
years,
then puts an upper limit $\abundchi<1$. This is shown in
Fig.~\ref{dm:fig}. We
see that {\em the whole} plane ($\mhalf,\mzero$)
becomes limited within a few hundred GeV.

The dominant effect is played here by the annihilation of the
neutralinos into
light fermion-antifermion pairs via the $t$-channel
exchange of the lightest sfermion(s); roughly
$\abundchi\propto m_{\tilde f}^4/\mchi^2$~\cite{dmreview,chiasdm},
although
including other final states affects the exact location of the bound.

It is interesting to explore how these bounds vary with different
choices of
the input
parameters. We find that at least for small $\tanb\lsim2$ one is able
to close
almost the whole plane ($\mhalf,\mzero$), and therefore the whole
SUSY
spectrum, from {\em above} for any
combinations of other parameters, except for the region of large
$Z$-pole
enhancement ($\mzero\gg\mhalf\simeq100\gev$). For larger values of
$\tanb$
sometimes the conditions of EWSB cannot be satisfied
in the regions of extreme $\mhalf$ or $\mzero$ and very close to such
regions
the LSP is of higgsino-type. In such, relatively rare, cases
one cannot close the plane ($\mhalf,\mzero$) from above completely.
It is remarkable that small $\tanb$ is
strongly favoured by the simple unification of the $b$- and
$\tau$-Yukawa
coupling unification, like in
$SU(5)$~\cite{bbop}.
It was recognized several years ago that, unlike in the SM, in the
MSSM the
$b$- and $\tau$-Yukawa running couplings meet at roughly the same
mass scale at
which the unification of the gauge couplings takes
place~\cite{bothunif}.
In Ref.~\cite{kkrwtwo} we have studied in detail the various
consequences of
adding this sensible, but rather specific to $SU(5)$-type GUTs,
assumption.

Can these upper bounds be improved?
It is worth stressing that the assumption that the age of the
Universe is at
least 10 billion years is actually a rather conservative one.
Many expect it to be no less than some 15 billion years which
translates to
$\abundchi\lsim0.25$ and much tighter bounds on the
parameters $\mhalf$ and $\mzero$. Another attractive hypothesis is
the
one of cosmic inflation which predicts $\Omega=1$ in which case most
of the matter in the Universe must most likely hide in the form of
DM.
Two scenarios have attracted a lot of attention. In the purely cold
DM
(CDM) scenario the neutralino would constitute most of DM in the
(flat)
Universe in which case the range $0.25\lsim\abundchi\lsim0.5$ would
be
favored. More recently (after COBE),
a mixed CDM+HDM picture (MDM) became more popular as it apparently
fits the astrophysical data
better than the pure CDM model.
In the mixed scenario one assumes about
30\% HDM (like light neutrinos with $m_\nu\simeq 6\ev$) and about
65\% CDM (bino-like neutralino), with baryons contributing
the remaining 5\% of the DM.
In this case the favored range for $\abundchi$ is approximately given
by
$0.16\lsim\abundchi\lsim0.33$. Both ranges are plotted in
Fig.~\ref{dm:fig}. It
is clear that their effect is to significantly reduce
the allowed parameter space from {\em both above and below}.
Consequently, the
allowed mass ranges of the various SUSY (and Higgs) particles become
much more
restricted and, unfortunately, typically beyond the reach of LEP~II
and the upgraded Tevatron.
More details can be found in
Refs.~\cite{kkrwone} and~\cite{kkrwtwo}.

\section{Conclusions}

I have shown that, in the framework of Constrained MSSM (which is the
MSSM with
a few well-motivated assumptions stemming from grand unifications),
one can
often limit the SUSY particle masses
below about 1-2\tev\ {\em by physical constraints alone}.
I have not used the ill-defined fine-tuning constraint {\em at all}.
It
certainly still makes sense to take it into account in expressing our
expectations as to where SUSY might be realized. But relatively
general
physical constraints now do not allow us to push SUSY into a
multi-\tev\ region
even if we wanted. Especially with the improving
knowledge of the top mass and the age of the Universe we soon will be
able to make a definite statement, based purely on physics criteria,
that
(minimal) low-energy SUSY is either
realized roughly below 1\tev\ or is not realized in Nature at all.

\section*{Acknowledgements}
I would like to thank Pierre Ramond for organizing a very stimulating
meeting. His warm reception especially contrasted with the freezing
weather
back in Michigan.

%%%%%%%%%%%%% begin refs %%%%%%%%%%%%%%%

%%%%%%%%%%%%% end of refs %%%%%%%%%%%%%%%


\begin{thebibliography}{99}
%
\bibitem{kkrwone} G.L.~Kane, C.~Kolda, L.~Roszkowski, and J.~Wells,
Michigan
preprint
UM-TH-93-24 (October 1993), Phys. Rev. {\bf D}, {\it in press}.
%
\bibitem{kkrwtwo}
C.~Kolda, L.~Roszkowski, J.~Wells, and G.L.~Kane, Michigan
preprint
UM-TH-94-03 (February 1994).
%
\bibitem{early}
P.~Langacker, in the {\em Proceedings of the PASCOS-90
Symposium}, eds.
P.~Nath and S.~Reucroft (World Scientific, Singapore, 1990);
P.~Langacker and M.-X.~Luo, \PRD{44}{91}{817};
J. Ellis, S. Kelley, and D.V.~Nanopoulos, \PLB{260}{91}{131};
U.~Amaldi, W.~de~Boer, and H.~F\"urstenau, \PLB{260}{91}{447};
F.~Anselmo, L.~Cifarelli, A.~Peterman, and A.~Zichichi, Nuovo Cim.
{\bf104A} (1991) 1817, and Nuovo Cim. {\bf105A} (1992) 581.
%
\bibitem{dgs}
See, \eg, M.~Dine, A.~Kagan, and S.~Samuel, \PLB{243}{90}{250}.
%
\bibitem{dmreview}
L.~Roszkowski, ``Supersymmetric Dark Matter - A Review'',
UM-TH-93-06, hep-ph/9302259 (February 1993), in the Proceedings of
the XXIII
Workshop {\em Properties of SUSY Particles}, Erice, Italy,
September~28 -
October~4, 1992, eds. L.~Cifarelli and V.A.~Khoze.
%
\bibitem{chiasdm}
L.~Roszkowski, \PLB{262}{91}{59}.
%
\bibitem{bbop}
V.~Barger, M.S.~Berger, P.~Ohmann, and R.J.N.~Phillips,
\PLB{314}{93}{351}.
%
\bibitem{bothunif}
H.~Arason, D.~Casta\~no, B.~Keszthelyi, S.~Mikaelian,
E.~Piard,
P.~Ramond, and B.~Wright, \PRL{67}{91}{2933}.
%
\end{thebibliography}
\end{document}